\documentclass[pra,aps,twocolumn,showpacs]{revtex4}

\usepackage{amssymb}
\usepackage{graphicx}

\usepackage{amsmath} 

\begin{document}

\title{Bounds on quantum process fidelity from minimum required number\\ of quantum state fidelity measurements}

\author{Jarom\'{i}r Fiur\'{a}\v{s}ek}
\affiliation{Department of Optics, Palack\'{y} University, 17. listopadu 1192/12, CZ-771 46 Olomouc, Czech Republic}

\author{Michal Sedl\'{a}k}
\affiliation{Department of Optics, Palack\'{y} University, 17. listopadu 1192/12, CZ-771 46 Olomouc, Czech Republic}

\begin{abstract}
To certify that an experimentally implemented quantum transformation is a certain unitary operation $U$ on a $d$-dimensional 
Hilbert space, it suffices to determine fidelities of output states for $d+1$ suitably chosen pure input states
[Reich \emph{et al.}, Phys. Rev. A \textbf{88}, 042309 (2013)]. The set of these $d+1$ probe states can consist of $d$ orthogonal states that form a basis and one additional state which is a balanced superposition 
of all $d$ basis states. 
Here we provide an analytical lower bound on quantum process fidelity for two-qubit quantum gates which results from the knowledge of average state fidelity for the  basis states and the fidelity 
of the superposition state. We compare this bound with the Hofmann bound that is based on knowledge of average state fidelities for two mutually unbiased bases. 
We also discuss possible extension of our findings to $N$-qubit operations.

\end{abstract}


\maketitle 

\section{Introduction}

Development and testing of advanced quantum information processing devices requires efficient 
 methods for their characterization. 
Full quantum process tomography \cite{Poyatos97,Chuang97,Jezek03} is suitable for small-scale devices such as two-qubit quantum gates. 
However, the number of measurements that need to be performed grows exponentially  with the number of qubits which makes this approach rather time-consuming 
and impractical for larger systems. Therefore, increasing attention has been paid in recent years to development of alternative less demanding 
techniques for assessment of the performance of quantum devices \cite{Hofmann05,Bendersky08,Gross10,Shabani11,Flammia11,Silva11,Reich13}. 
Typically, the device is probed by various input states, and measurements 
are performed on the output states. In this context, one may ask what is the minimum necessary number of input states to certify that the 
implemented quantum operation is a certain unitary operation $U$ on a $d$-dimensional Hilbert space.
 Very recently, Reich \emph{et al.} showed that $d+1$ pure input probe states are sufficient for this purpose \cite{Reich13}. 
 A suitable set of these states consists of $d$ basis states together with one additional  
state  which is a superposition of all the basis states. 
If the output state fidelities for these $d+1$ input states are all equal to $1$, then the implemented quantum operation 
 must be exactly the target unitary operation $U$ \cite{Reich13}. 

An appealing feature of this approach is that it requires only $d+1$ input probe states in comparison to $d^2$ probe states necessary for full quantum process tomography. 
These results are similar in spirit to the earlier findings by Hofmann \cite{Hofmann05}, who derived an analytical lower bound on quantum process fidelity 
$F_\chi$ in terms of average state fidelities $F$ and $F'$ for two mutually unbiased bases, $F_\chi \geq F+F'-1$. Clearly, if $F'=F=1$ then also $F_\chi=1$. 
The Hofmann bound was successfully applied to characterization of several experimentally implemented two-qubit and three-qubit quantum operations \cite{Okamoto05,Bao07,Clark09,Gao10,Zhou11,Lanyon11,Micuda13}. 
Finding a similar analytical lower bound on quantum process fidelity also for the scheme with the minimum number of $d+1$ pure probe states  turns out to be more difficult 
due to reduced symmetry. In Ref. \cite{Reich13}, this problem was studied numerically and although an expression quantifying the quantum gate performance as a function of state fidelities was proposed, 
it was noted that it can lead to underestimation of gate error in certain cases.

Here we derive an exact lower bound on quantum process fidelity of two-qubit operations based on knowledge of the average state fidelity $F$ 
for certain basis $|j\rangle$ and state fidelity $G$ for a state $|s\rangle$ which is a balanced superposition of all the basis states $|j\rangle$.
We compare this bound with the Hofmann bound whose determination requires measurement of state fidelities for two bases and we find that the new bound is typically 
much weaker than the Hofmann bound. Therefore, the number of probe states can be reduced from $2d$ to $d+1$ only at a cost of potentially much less precise device characterization.

The rest of the paper is organized as follows. In Section II we provide an explicit construction of a two-qubit quantum operation (a trace-preserving completely positive map) 
which for given state fidelities $F$ and $G$ achieves minimum quantum process fidelity. This fidelity thus provides
 a lower bound on process fidelity of any operation achieving state fidelities $F$ and $G$. Analytical proof of this bound is provided in Section III and extension 
 of our construction to $N$-qubit operations is proposed in Section IV. Although we do not provide any rigorous optimality proof for the $N$-qubit case, 
 our construction nevertheless illustrates that the gap between the fidelity bound and the true fidelity will typically increase fast with the growing number of qubits. 
 Finally, Section V contains a brief summary and conclusions.

\section{Two-qubit operations}

Let $|0\rangle$ and $|1\rangle$ denote the computational basis states of a single qubit and define superposition states $|\pm\rangle=\frac{1}{\sqrt{2}}(|0\rangle\pm|1\rangle)$.
We would like to determine a lower bound on fidelity of a two-qubit quantum operation $\mathcal{E}$ with unitary operation $U$ 
provided that we know the average output state fidelity $F$ for the computational basis $|jk\rangle$, where $j,k\in\{0,1\}$, 
and also output state fidelity $G$ for input state $|\!+\!+\rangle$ which is a balanced superposition of all computational basis states, 
\begin{equation}
|\!+\! +\rangle=\frac{1}{2}\left(|00\rangle +|01\rangle+|10\rangle+|11\rangle \right).
\end{equation}
According to the Choi-Jamiolkowski isomorphism \cite{Choi75, Jamiolkowski72}, any quantum operation $\mathcal{E}$ can be represented by a positive semidefinite operator $\chi$ on the 
tensor product of input and output Hilbert spaces. Given input state $\rho_{\mathrm{in}}$, the output state can be calculated according to
\begin{equation}
\rho_{\mathrm{out}}= \mathrm{Tr}_{\mathrm{in}}[\rho_{\mathrm{in}}^T \otimes \mathbb{I}_{\mathrm{out}} \, \chi], 
\end{equation}
where $\mathrm{Tr}_{\mathrm{in}}$ denotes the partial trace over the input Hilbert space, $\mathbb{I}$ denotes the identity operator, 
and $T$ stands for transposition in the computational basis. We shall consider deterministic operations described by trace-preserving maps. 
The trace-preservation condition can be expressed as
\begin{equation}
\mathrm{Tr}_{\mathrm{out}}[\chi]=\mathbb{I}_{\mathrm{in}},
\label{tpcondition}
\end{equation}
and it guarantees that $\mathrm{Tr}[\rho_{\mathrm{out}}]=\mathrm{Tr}[\rho_{\mathrm{in}}]$ for arbitrary $\rho_{\mathrm{in}}$. 
In this formalism, a unitary operation $U$ is isomorphic to a pure maximally entangled state,
\begin{equation}
\chi_U= 4|\chi_U\rangle \langle \chi_U| , 
\end{equation}
where $|\chi_U\rangle =(\mathbb{I}_{\mathrm{in}}\otimes U)\, |\Phi_2^{+}\rangle $,
\begin{equation}
|\Phi_2^{+}\rangle= \frac{1}{2}\sum_{j,k=0}^1 |jk\rangle \otimes |jk\rangle 
\end{equation}
is a maximally entangled state between qubits in the input Hilbert space and output Hilbert space, and the factor $4$ ensures correct 
normalization of $\chi_U$ as implied by the trace-preservation condition (\ref{tpcondition}).

The average output state fidelity for computational basis is defined as
\begin{equation}
F= \frac{1}{4} \sum_{j,k=0}^1 \langle jk|U^\dagger \rho_{\mathrm{out}}^{jk} U|jk\rangle,
\label{Fdefinition}
\end{equation}
 where $U|jk\rangle$ is a pure output state that would be generated by the unitary $U$ and 
 $\rho_{\mathrm{out}}^{jk}= \mathrm{Tr}_{\mathrm{in}}[|jk\rangle\langle jk| \otimes \mathbb{I}_{\mathrm{out}} \,\chi]$ is the output state
 produced by the actually implemented quatum operation $\chi$. On inserting the formula for $\rho_{\mathrm{out}}^{jk}$ into Eq. (\ref{Fdefinition}), we obtain \cite{Micuda13}
 \begin{equation}
 F=\mathrm{Tr}[(\mathbb{I}\otimes U) R_F(\mathbb{I}\otimes U^\dagger) \chi   ],
 \label{Fmatrixform}
 \end{equation}
 where 
 \begin{equation}
 R_F=\frac{1}{4} \sum_{j,k=0}^1 |jk\rangle \langle jk| \otimes |jk\rangle \langle jk|.
 \end{equation}
 Similarly, fidelity  of the output state $\rho_{\mathrm{out}}^{+}$ obtained from input state $|\!+\!+\rangle$ is defined as
 \begin{equation}
 G=\langle +\!+\!|U^\dagger \rho_{\mathrm{out}}^{+} U|\!+\!+\rangle.
 \label{Gdefinition}
 \end{equation}
 We can express this fidelity as 
 \begin{equation}
 G= \mathrm{Tr}[(\mathbb{I}\otimes U) R_G (\mathbb{I}\otimes U^\dagger) \chi   ],
 \label{Gmatrixform}
 \end{equation}
 where
 \begin{equation}
 R_G= |\!+\!+\rangle\langle +\!+\!| \otimes |\!+\!+\rangle \langle +\!+\!|.
 \end{equation}
 Our goal is to determine a lower bound on quantum process fidelity 
 \begin{equation}
 F_{\chi}= \frac{\mathrm{Tr}[\chi_U \chi]}{\mathrm{Tr[\chi_U]\mathrm{Tr}[\chi]}} 
 \label{Fchidefinition}
 \end{equation}
 from the knowledge of state fidelities $F$ and $G$. For a two-qubit unitary operation $U$ we explicitly have
 \begin{equation}
 F_{\chi}=\frac{1}{4} \langle \chi_U| \chi |\chi_U\rangle.
 \label{FchiU}
 \end{equation}

 We shall proceed by constructing a particular quantum operation $\tilde{\chi}$ that achieves the state fidelities $F$ and $G$. We then prove that the quantum process fidelity  of this particular
 operation provides a lower bound on $F_\chi $, i.e. it represents the lowest possible value of $F_\chi$  consistent with $F$ and $G$. 
 Our \emph{ansatz} for the quantum operation $\tilde{\chi}$ reads
 \begin{equation}
 \tilde{\chi} =(\mathbb{I}\otimes U) \,\tilde{\chi}_S \,(\mathbb{I}\otimes U^\dagger ),  
 \label{chitilde}
 \end{equation}
 where  
 \begin{equation}
 \tilde{\chi}_S=\sum_{m=0}^3 |\chi_m\rangle \langle \chi_m| 
 \label{chitildeS}
 \end{equation}
 and 
\begin{eqnarray*}
 |\chi_0\rangle&=& a\,Z_{00}|\Phi_2^{+}\rangle + b\,|\!+\!+\rangle |\!+\!+\rangle, \\[2mm]
 |\chi_1\rangle&=& c\,Z_{01}|\Phi_2^{+}\rangle + d\,|\!+\!+\rangle |\!+-\rangle, \\[2mm]
 |\chi_2\rangle&=& c\,Z_{10}|\Phi_2^{+}\rangle + d\,|\!+\!+\rangle |-\!+\rangle, \\[2mm]
 |\chi_3\rangle&=& c\,Z_{11}|\Phi_2^{+}\rangle + d\,|\!+\!+\rangle |-\!-\rangle. 
 \label{chim}
 \end{eqnarray*}
 Here $Z_{jk}=\mathbb{I}_{\mathrm{in}}\otimes \sigma_Z^j \otimes \sigma_Z^k$, the Pauli matrix $\sigma_Z$ is defined as $\sigma_Z=|0\rangle\langle 0|-|1\rangle\langle 1|$, $\sigma_Z^0=\mathbb{I}$, and $\sigma_Z^1=\sigma_Z$.
 The complete positivity condition $\tilde{\chi} \geq 0$ is satisfied by construction. On inserting the ansatz (\ref{chitilde}) into Eq. (\ref{tpcondition}), we find after some algebra
 that the trace-preservation condition is equivalent to the following constraints:
 \begin{eqnarray}
 a^2+3c^2&=&4, \nonumber \\
 b^2+ab+3d^2+3cd&=&0.
 \label{tpabcd}
 \end{eqnarray}
  The output state fidelities $F$ and $G$ can be determined by inserting $\tilde{\chi}$ into Eqs. (\ref{Fmatrixform}) and (\ref{Gmatrixform}), respectively. We get
  \begin{equation}
  F= \frac{1}{16}(2a+b)^2+\frac{3}{16}(2c+d)^2 ,
  \label{Fabcd}
  \end{equation} 
  and
 \begin{equation}
 G=\frac{(2b+a)^2}{4}.
 \label{Gabcd}
 \end{equation}
  Finally, with the help of Eq. (\ref{FchiU}) we can also calculate the quantum process fidelity of operation $\tilde{\chi}$ with unitary operation $U$,
 \begin{equation}
 \tilde{F}_\chi=\frac{(2a+b)^2}{16}.
 \label{Fchiabcd}
 \end{equation}
 
  \begin{figure}[!b!]
\centerline{\includegraphics[width=0.95\linewidth]{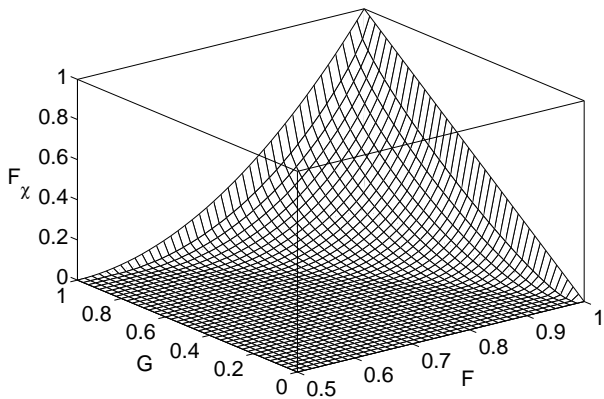}}
\caption{Lower bound on quantum process fidelity $F_{\chi}$ of a two-qubit quantum operation is plotted as a function of state fidelities $F$ and $G$.}
\end{figure}

 The system of equations (\ref{tpabcd}), (\ref{Fabcd}), and (\ref{Gabcd}) can be solved and the parameters $a$, $b$, $c$, and $d$ can be expressed in terms of the output state fidelities,
 \begin{eqnarray}
 a &=& \frac{2}{3} \left[(8F-5)\sqrt{G}-4\sqrt{(1-F)(4F-1)(1-G)} \right],   \nonumber \\
 b &=& \sqrt{G}-\frac{a}{2},  \nonumber \\
 c&=&\sqrt{\frac{4-a^2}{3}}, \nonumber \\
 d &=& \sqrt{\frac{1-G}{3}} - \frac{1}{2}\sqrt{\frac{4-a^2}{3}}.
 \label{aformula}
 \end{eqnarray}
 On inserting the expressions for $a$ and $b$ into Eq. (\ref{Fchiabcd}) we finally arrive at a formula for $\tilde{F}_\chi$ as a function of $F$ and $G$,
 \begin{equation}
 \tilde{F}_\chi=\left[(2F-1)\sqrt{G}-\sqrt{(4F-1)(1-F)}\sqrt{1-G} \right]^2.
 \label{Fchibound}
 \end{equation}
 We prove in the next Section that $\tilde{F}_\chi$ provides a lower bound on the quantum process fidelity $F_\chi$ provided that 
 \begin{equation}
 F\geq F_{\mathrm{th}},
 \end{equation}
  where 
 \begin{equation}
 F_{\mathrm{th}}=\frac{1}{8}\left(5 - G + \sqrt{9 - 10 G + G^2}\right).
 \end{equation}
 The fidelity threshold $F_{\mathrm{th}}$ is determined by the condition that $\tilde{F}_\chi=0$ when $F=F_{\mathrm{th}}$. 
 If  $F<F_{\mathrm{th}}$ then the state fidelities $F$ and $G$ are consistent with $F_\chi=0$. In Fig. 1 
 we plot the lower bound on $F_\chi$ as a function of $F$ and $G$. We emphasize that our calculation is valid for arbitrary 
 two-qubit unitary operation $U$. Therefore, the bound $\tilde{F}_\chi$ is also universally valid.

 \begin{figure}[!t!]
\centerline{\includegraphics[width=0.9\linewidth]{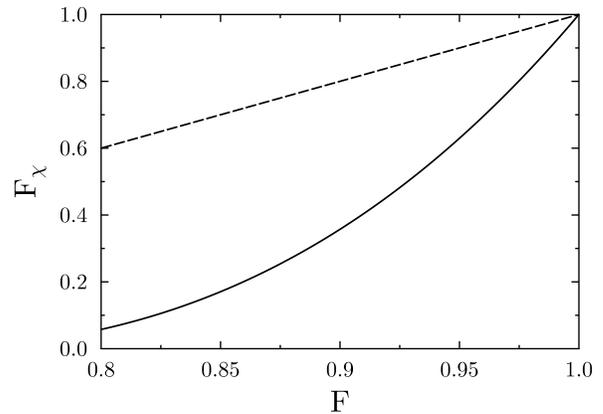}}
\caption{Lower bound on quantum process fidelity $F_\chi$ determined from the knowledge of state fidelities $F$ and $G$ (solid line) 
and the Hofmann lower bound on quantum process fidelity (dashed line) are plotted for two-qubit quantum operations assuming $G=F'=F$.}
\end{figure}

It is instructive to compare this lower bound with the Hofmann bound $F_\chi \geq F+F'-1$ which is based on the knowledge 
of average output state fidelities $F$ and $F'$ for two mutually unbiased bases. To make such comparison possible, 
we shall assume that the average output state fidelity $F'$ for basis which is  mutually unbiased with the computational basis 
and which contains state $|\!+\!+\rangle$ is equal to the fidelity $G$. 
Let us consider high-fidelity operation, $F=1-\epsilon$ and $G=1-\delta$, where $\epsilon,\delta \ll 1$. If we keep only terms up to linear in $\delta$ and $\epsilon$, we obtain
\begin{equation}
F_\chi \geq 1-\epsilon -\delta
\end{equation}
for the original Hofmann bound, and 
\begin{equation}
F_\chi \gtrsim 1-4\epsilon -\delta -2\sqrt{3} \sqrt{\epsilon \delta}
\end{equation}
for the bound (\ref{Fchibound}). We can see that the Hofmann bound is higher than the bound (\ref{Fchibound}) and the gap increases with decreasing fidelity $F$. This is illustrated in Fig. 2 where we plot both
bounds as a function of $F$ assuming that $F=G$. To provide some scale we note that, for instance, fidelity  of two-qubit {\sc cnot} gate with the identity operation reads $0.25$.
We can thus conclude that the state fidelities $F$ and $G$ have to be very high to obtain a meaningful and nontrivial bound on process fidelity from Eq. (\ref{Fchibound}).

\section{Optimality Proof}

Here we prove that the two-qubit quantum operation (\ref{chitilde}) constructed in the previous section 
exhibits minimum quantum process fidelity $F_\chi$ compatible with $F$ and $G$. The proof is based on the techniques from 
semidefinite programming \cite{Vandenberghe96,Audenaert02}. We define an operator
\begin{equation}
M=\frac{1}{4}|\Phi_2^+\rangle \langle \Phi_2^{+}|+ x R_F +w R_G + 
y\mathbb{I} + z |\!+\!+\rangle\langle +\!+\!|\otimes \mathbb{I}_{\mathrm{out}}.
\label{Mdefinition}
\end{equation}
The parameters appearing in definition of $M$ can be interpreted
as Lagrange multipliers that account for a fixed value of $F$ and $G$, and for the trace preservation condition (\ref{tpcondition}). 
Suppose that we choose the Lagrange multipliers such that
\begin{equation}
(\mathbb{I}\otimes U)  M  (\mathbb{I}\otimes U^\dagger) \,  \tilde{\chi} =0,
\label{Mslackness}
\end{equation}
and
\begin{equation}
M \geq 0.
\label{Mpositivity}
\end{equation}
Note that the condition (\ref{Mslackness}) can be equivalently expressed as 
\begin{equation}
M \tilde{\chi}_S=0,
\label{MchiS}
\end{equation}
hence it is independent on the unitary operation $U$. It follows from the inequality $M\geq 0$ that 
\begin{equation}
\mathrm{Tr}\left[(\mathbb{I}\otimes U)  M  (\mathbb{I}\otimes U^\dagger) \,  \chi \right] \geq 0,
\label{Mchipositive}
\end{equation}
for arbitrary trace preserving map $\chi$, because trace of product of two positive semidefinite operators is nonnegative. 
On inserting the expression (\ref{Mdefinition}) into Eq. (\ref{Mchipositive}) we obtain
\begin{equation}
F_\chi +xF+wG+4y+z\geq 0.
\end{equation}
By taking trace of Eq. (\ref{MchiS}) we find that $\tilde{F}_\chi= -[xF+wG+4y+z]$. 
Therefore, the positive semidefiniteness of $M$ together with the condition (\ref{MchiS}) implies a lower bound on quantum process fidelity,
\begin{equation}
F_\chi \geq \tilde{F}_\chi.
\end{equation}
In what follows we provide explicit formulas for the Lagrange multipliers and prove that $M \geq 0$. 

The Lagrange multipliers can be determined from Eq. (\ref{MchiS}) which is equivalent to $4$ conditions $M|\chi_m\rangle=0$, $m=0,1,2,3$. 
This provides a system of $4$ linear equations for the $4$ unknown parameters $x$, $w$, $y$, and $z$,
\begin{eqnarray}
2a(1+x+4y)+b(1+x)&=&0,\nonumber  \\
a(w+z)+2b(w+y+z)&=&0, \nonumber  \\
2c(x+4y)+dx&=&0, \\
cz+2d(y+z)&=&0. \nonumber 
\end{eqnarray}
If we solve this system of equations and express $b$, $c$, and $d$ as functions of $a$ and $G$, c.f. Eq. (\ref{aformula}),
we obtain
\begin{eqnarray*}
x&=&\frac{\sqrt{4-a^2}\,(3a+2\sqrt{G})}{2a\sqrt{1-G}-2\sqrt{G(4-a^2)}}, \\
w&=&-\left(3\sqrt{4-a^2}+2\sqrt{1-G}\right)\frac{3a+2\sqrt{G}}{64\sqrt{G(1-G)}}, \\
y&=&\frac{1}{32}\left(3a+2\sqrt{G}\right)\frac{3\sqrt{4-a^2}+2\sqrt{1-G}}{\sqrt{G(4-a^2)}-a\sqrt{1-G}}, \\
z&=& \frac{\sqrt{4-a^2}-2\sqrt{1-G}}{2\sqrt{1-G}}\,y.
\end{eqnarray*}
Since the condition $M\tilde{\chi}_S=0$ is satisfied by construction, it remains to prove that $M \geq 0$.
The eigenvalues of operator $M$ are listed below,
\begin{eqnarray}
 \lambda_1=y,\qquad  \qquad \qquad \quad  \nonumber \\
\lambda_{2}=\frac{1}{8}(A-\sqrt{B}), \qquad \lambda_{3}=\frac{1}{8}(A+\sqrt{B}), \\
\lambda_4=\frac{1}{8}(C-\sqrt{D}),  \qquad
\lambda_5=\frac{1}{8}(C+\sqrt{D}). \nonumber
\end{eqnarray}
Here
\begin{equation}
A=x+8y+4z, \qquad B=x^2-4xz+16z^2,
\end{equation}
\begin{equation}
C=1+4w+x+8y+4z,
\end{equation}
and
\begin{equation}
D=1+16w^2+2x+x^2-4w(1+x-8z)-4z-4xz+16z^2.
\end{equation}
The eigenvalue $\lambda_1$ is $8$-fold degenerate and the eigenvalues $\lambda_2$ and $\lambda_3$ are each $3$-fold degenerate.
One can verify by direct calculation that $A^2=B$ and $C^2=D$. The operator $M$ is thus positive semidefinite if
$y \geq 0$, $A \geq 0$, and $C\geq 0$. After some algebra we find that 
\begin{equation}
A=\frac{3a+2\sqrt{G}}{16\sqrt{1-G}}\frac{16-3a^2-4G}{\sqrt{G(4-a^2)}-a\sqrt{1-G}},
\end{equation}
and
\begin{equation}
C=\frac{3a^2+4G}{16\sqrt{G}} \frac{3\sqrt{4-a^2}+2\sqrt{1-G}}{\sqrt{G(4-a^2)}-a\sqrt{1-G}}.
\end{equation}

We shall first derive several useful auxiliary inequalities. Since $F \leq 1$, it follows from  Eq. (\ref{aformula}) that
\begin{equation}
a \leq 2\sqrt{G}.
\label{amax}
\end{equation}
Assuming a fixed $G$, $a$ is a monotonically increasing function of $F$ in the interval $F \in[F_{\mathrm{th}},1]$. Parameter $a$ as a function of $F$
exhibits a single local minimum at $F_0=(5 - 3\sqrt{G})/8$ and $F_0 \leq F_{\mathrm{th}}$ for all $G\in [0,1]$. Minimum value of $a$ for a fixed $G$
in the interval $F \in[F_{\mathrm{th}},1]$ is thus achieved at $F=F_{\mathrm{th}}$ and we have
\begin{equation}
a \geq -\frac{2\sqrt{G}}{3}.
\label{amin}
\end{equation}
Recall that we have to restrict ourrselves to $F \geq F_{\mathrm{th}}$ because 
for $F<F_{\mathrm{th}}$ the operation (\ref{chitilde}) does not provide the minimum quantum process fidelity compatible with given $F$ and $G$. 
Inequalities (\ref{amax}) and (\ref{amin}) imply that $a^2 \leq 4G$, hence
\begin{equation}
3a^2+4G \leq 16G \leq 16.
\label{asquared}
\end{equation}
Furthermore, it follows from the inequality $a\leq 2\sqrt{G}$ that 
\begin{equation}
\sqrt{G(4-a^2)}-a\sqrt{1-G} \geq 0.
\label{Ginequality}
\end{equation}
The inequalities $y \geq 0$, $A \geq 0$ and $C \geq 0$ now directly follow from the inequalities (\ref{amin}), (\ref{asquared}) and (\ref{Ginequality}).
This proves that $M \geq 0$ in the entire domain $F \geq F_{\mathrm{th}}$.

For the sake of completeness, we also explicitly show that if $F < F_{\mathrm{th}}$ then the lower bound on quantum process fidelity reads $\tilde{F}_\chi=0$. It is sufficient to find quantum operations
with $F_{\chi}=0$ for three boundary points $F=0$ and $G=1$, $F=1$ and $G=0$, and $F=0$ and $G=0$. At any point in the area $F < F_{\mathrm{th}}$ a quantum operation with given $F$, $G$ and $\tilde{F}=0$
can then be constructed as a mixture of these three operations and operations (\ref{chitilde}) corresponding to the boundary line $F=F_\mathrm{th}$, where $\tilde{F}_\chi=0$.
Fidelities $F=0$, $G=1$, and $F_\chi=0$ can be achieved by a unitary operation $UX$,
where $X=\sigma_X\otimes \sigma_X$ performs a bit flip on each qubit, $\sigma_X=|0\rangle\langle 1|+|1\rangle\langle 0|$. 
Similarly, $F=1$, $G=0$, and $F_\chi=0$ is achieved by operation $UZ$, where phase flips are inserted before $U$, $Z=\sigma_Z\otimes \sigma_Z$. 
Finally, if we combine both bit flips and phase flips we obtain unitary operation $UZX$ which exhibits $F=0$, $G=0$, and $F_\chi=0$.

\section{N-qubit operations}

In this section we generalize the construction of quantum operation (\ref{chitilde}) to $N$-qubit operations.
 Note that in contrast to the two-qubit operations we do not prove that the resulting expression for quantum process fidelity is a lower bound on $F_\chi$. Nevertheless, this approach allows us to investigate the
 scaling of the bound with the number of qubits. In particular, we shall show that under reasonable assumptions the gap between the Hofmann bound and the bound determined by average state fidelity $F$
 and fidelity of a single superposition state $G$ grows exponentially with the number of qubits $N$. 
 
 We shall label the $N$-qubit computational basis states by an integer $j$, $0 \leq j \leq 2^N-1$. We define $|j\rangle=|j_1\rangle|j_2\rangle \cdots|j_N\rangle$, where $j_k$ denotes 
 $k$th digit of binary representation of integer $j$. With this notation at hand, we can define the average state fidelity $F$ in the computational basis, 
 \begin{equation}
 F= \frac{1}{2^N}\sum_{j=0}^{2^N-1} \mathrm{Tr}\left[|j\rangle\langle j|\otimes U|j\rangle\langle j|U^\dagger \, \chi\right].
 \end{equation}
 Similarly, we can define fidelity $G$ of output state obtained from the input superposition state 
 \begin{equation}
 |s\rangle= \frac{1}{\sqrt{2^{N}}}\sum_{j=0}^{2^N-1}|j\rangle,
 \end{equation}
 and we have  $G= \mathrm{Tr}[|s\rangle\langle s|\otimes U|s\rangle\langle s|U^\dagger \, \chi].$
 
 Analogically to the  two-qubit case we construct an $N$-qubit quantum operation, 
 \begin{equation}
 \tilde{\chi}= (\mathbb{I}\otimes U) \tilde{\chi}_S (\mathbb{I}\otimes U^\dagger),
 \end{equation}
  where 
  \begin{equation}
  \tilde{\chi}_S=\sum_{j=0}^{2^N-1} |\chi_j\rangle\langle \chi_j|.
  \end{equation}
 The constituents $|\chi_j\rangle$ can be expressed as superpositions of maximally entangled states and product states of qubits in intput and output Hilbert spaces. For $j=0$ we have
 \begin{equation}
 |\chi_0\rangle= a |\Phi_N^{+}\rangle +b |s\rangle|s\rangle,
 \end{equation}
 while for $j \geq 1$ we define
 \begin{equation}
 |\chi_j\rangle= \mathbb{I}\otimes V_{j}  \left(c\,|\Phi_N^{+}\rangle +d\, |s\rangle|s\rangle\right).
 \end{equation}
 Here 
 \begin{equation}
 |\Phi_N^{+}\rangle=\frac{1}{\sqrt{2^N}}\sum_{j=0}^{2^N-1}|j\rangle|j\rangle
 \end{equation}
 is a maximally entangled state of $2N$ qubits, and the $N$-qubit unitary operators $V_j$ are defined as products of $\sigma_Z$ and identity operators,
 \begin{equation}
 V_j=\bigotimes_{k=1}^N  \sigma_Z^{j_k}.
 \end{equation}

 The trace preservation condition $\mathrm{Tr}_{\mathrm{out}}[\tilde{\chi}]=\mathbb{I}_{\mathrm{in}}$ is equivalent to 
 \begin{eqnarray}
& a^2+(2^N-1)c^2=2^N, & \nonumber \\
 & b^2+2^{1-N/2}ab+(2^N-1)\left(d^2+2^{1-N/2}cd\right)=0, &
 \label{tpNqubit}
 \end{eqnarray}
and the fidelities $F$ and $G$ can be expressed as 
 \begin{eqnarray}
 F&=&\frac{1}{2^N}\left(a+\frac{b}{2^{N/2}}\right)^2+\left(1-\frac{1}{2^N}\right)\left(c+\frac{d}{2^{N/2}}\right)^2, \nonumber \\
  G&=&\left(\frac{a}{2^{N/2}}+b\right)^2, \nonumber \\
 \label{FGNqubit}
 \end{eqnarray}
Recall that  $N$-qubit unitary operation $U$ is represented by $\chi_{U}=2^N |\chi_U\rangle\langle \chi_U|$, where $|\chi_U\rangle= \mathbb{I}\otimes U |\Phi_N^{+} \rangle$.
If we insert the $N$-qubit operations $\chi_U$ and $\tilde{\chi}$ into the formula for quantum process fidelity, Eq. (\ref{Fchidefinition}), we obtain
 \begin{equation}
 \tilde{F}_\chi= \frac{1}{2^N}\left(a+\frac{b}{2^{N/2}}\right)^2.
 \end{equation}
Note, that for $N=2$ we recover the formulas derived in Section II. With the help of the expressions (\ref{tpNqubit}) and (\ref{FGNqubit}) 
we obtain after some algebra formula for the quantum process fidelity $\tilde{F}_\chi$
as a function of the state fidelities $F$ and $G$,
\begin{eqnarray}
\tilde{F}_\chi&=&\left\{\left[1-(1-F)2^{N-1}\right]\sqrt{G} \right. \nonumber \\
& & \left. -\sqrt{(1-F)(1-G)}\sqrt{2^N-1-(1-F)2^{2N-2}}\right\}^2. \nonumber \\
\label{FchiNqubit}
\end{eqnarray} 
 The quantum process fidelity $\tilde{F}_{\chi}$ vanishes for $F=F_{\mathrm{th}}$, where
\begin{equation}
F_{\mathrm{th}}=1-\frac{1}{2^{N-1}}+\frac{1-G}{2^{2N-1}}+\frac{2}{2^{2N}}\sqrt{(1-G)\left[(2^N-1)^2-G\right]}.
\end{equation}
The lower bound on quantum process fidelity will thus certainly be zero when $F \leq F_{\mathrm{th}}$. It follows that with increasing number of qubits $N$ the fidelity $F$ has to be exponentially close to $1$, 
 $F \geq 1-2^{1-N}$, in order to obtain a nontrivial lower bound on $F_\chi$. Assuming fixed state fidelities $F$ and $G$, the bound on $F_\chi$ determined by these fidelities will thus quickly 
 become very low with growing $N$.  This should be contrasted with the Hofmann bound $F_\chi \geq F+F'-1$ whose form does not depend on the number of qubits $N$.
 We stress that for $N>2$ we did not prove that $\tilde{F}_\chi$  is the ultimate lower bound on $F_\chi$ for given $F$ and $G$. However, any such ultimate bound can only be smaller than $\tilde{F}_\chi$. 
 Therefore, our conclusions concerning the scaling with $N$ would hold even if $\tilde{F}_\chi$ would not be the ultimate lower bound on $F_\chi$.

 \section{Conclusions}
 In summary, we have derived a lower bound on fidelity of two-qubit quantum gates imposed by the knowledge of average state fidelity $F$ for one basis 
 and knowledge of state fidelity $G$ for one additional balanced superposition state. In our calculations we have explicitly considered the computational basis 
 but in practice this basis may be arbitrary, because any basis transformation can be included into the unitary operation $U$. We have seen that the quantum gate characterization 
 with the minimum number of pure probe states would generally yield rather low bound on process fidelity. This bound is particularly
 sensitive to the value of the average state fidelity $F$. In any potential experimental application of this technique one should therefore choose
 the basis for which one expects the best performance. At a cost of doubling the number of probe states
 one could instead determine the original Hofmann bound that will be typically significantly higher. Therefore, characterization by the minimum number 
 of probe states would be suitable in situations where the gate exhibits high fidelity and where measurements for the basis states are easy to implement 
 while measurements for the superposition states are very difficult and demanding.

  \acknowledgments
 J.F. acknowledges support by the Czech Science Foundation (Project No. 13-20319S). 
 M.S. acknowledges support by the Operational Program Education for Competitiveness - European Social Fund 
 (project No. CZ.1.07/2.3.00/30.0004) of the Ministry of Education, Youth and Sports of the Czech Republic.

\end{document}